\documentclass [a4paper,twoside,11pt] {article}
\usepackage {a4wide,amsmath,amssymb,latexsym,longtable}
\usepackage {a4wide,amsmath,amssymb,latexsym,graphicx}
\usepackage{rotating}

\newcommand{\mrm}{\mathrm}
\newcommand{\mbb}{\mathbb}

\newcommand{\bool}[1]{\{ 0,1\}^{#1}}

\newcommand{\und}{\underbrace}

\newcommand{\boite}[1]{B_0  B_1  \cdots B_{#1}}
\newcommand{\N}{\mathbb{N}}
\newcommand{\moins}{\dot{-}}
\newcommand{\ext}{\mathrm{ext} }

\newcommand{\size}{\ensuremath{\mathsf{SIZE}}}
\newcommand{\Rplus}{\ensuremath{\mathbb{R_{+}}}}

\newcommand{\existsio}{\exists^{\infty}}
\newcommand{\forallio}{\forall^{\infty}}

\newcommand{\dom}{\mathrm{dom}}

\newcommand{\pos}{\mathrm{pos}}
\newcommand{\bit}{\mathrm{bit}}

\newcommand{\p}{\ensuremath{\mathsf{P}}}
\newcommand{\np}{\ensuremath{\mathsf{NP}}}
\newcommand{\qp}{\ensuremath{\mathsf{QUASIPOLY}}}
\newcommand{\subexp}{\ensuremath{\mathsf{SUBEXP}}}
\newcommand{\rp}{\ensuremath{\mathsf{RP}}}
\newcommand{\bpp}{\ensuremath{\mathsf{BPP}}}

\newcommand{\zpp}{\ensuremath{\mathsf{ZPP}}}
\newcommand{\pspace}{\ensuremath{\mathsf{PSPACE}}}

\newcommand{\e}{\ensuremath{\mathsf{E}}}
\newcommand{\espace}{\ensuremath{\mathsf{ESPACE}}}
\newcommand{\expc}{\ensuremath{\mathsf{EXP}}}
\newcommand{\dtime}{\ensuremath{\mathsf{DTIME}}}
\newcommand{\bptime}{\ensuremath{\mathsf{BPTIME}}}

\newcommand{\dspace}{\ensuremath{\mathsf{DSPACE}}}
\newcommand{\sparse}{\ensuremath{\mathsf{SPARSE}}}
\newcommand{\rec}{\ensuremath{\mathsf{REC}}}
\newcommand{\qplin}{\ensuremath{\mathsf{QUASIPOLY_{lin}}}}

\newcommand{\spp}{\ensuremath{\mathsf{SPP}}}

\def\squareforqed{\hbox{\rlap{$\sqcap$}$\sqcup$}}
\def\qed{\ifmmode\squareforqed\else{\unskip\nobreak\hfil
\penalty50\hskip1em\null\nobreak\hfil\squareforqed
\parfillskip=0pt\finalhyphendemerits=0\endgraf}\fi}

\newenvironment{proof}{  \noindent\textit{Proof}.}{} 
\newtheorem{definition}{Definition}[section]
\newtheorem{theorem}{Theorem}[section]
\newtheorem{lemma}{Lemma}[section]
\newtheorem{corollary}{Corollary}[section]

\newtheorem{remark}{Remark}[section]

\newcounter{myenumi}
  {\begin{list}{\textup{(\arabic{myenumi}) }}{\usecounter{myenumi}%
    \setlength{\labelsep}{0pt}\setlength{\leftmargin}{0pt}%
    \setlength{\labelwidth}{0pt}%
    \setlength{\listparindent}{0pt}}}
  {\end{list}}

\newcounter{myenumii}
  {\begin{list}{\textup{(\arabic{myenumi}.\arabic{myenumii}) }} {\usecounter{myenumii}%
    \setlength{\labelsep}{0pt}\setlength{\leftmargin}{0pt}%
    \setlength{\labelwidth}{0pt}%
    \setlength{\listparindent}{0pt}}}
  {\end{list}}

\newcommand{\rpos}{\mathrm{rpos}}
\newcommand{\Tc}{\mathrm{T}}
\newcommand{\Sc}{\mathrm{S}}
\newcommand{\BPc}{\mathrm{BP}}
\newcommand{\Small}{\mathrm{SMALL}}
\newcommand{\cla}{\mathsf{C}}
\newcommand{\loc}{\mathrm{loc}}

\title{Baire Categories on Small Complexity Classes and Meager-Comeager Laws}
\author{Philippe Moser\footnote{Address: Email: mosersan@gmail.com}}
\date{}

\begin{document}
\maketitle


\begin{abstract}
	We introduce two resource-bounded Baire category notions on small complexity
        classes such as \p , \qp , \subexp\ and $\pspace$ and on probabilistic classes such as \bpp ,
	which differ  on how the corresponding finite extension strategies are computed.
        We give an alternative characterization of small sets via resource-bounded
        Banach-Mazur games.
        As an application of the first notion, we show that for almost every language   $A$ (i.e. all except a meager class)
	computable in subexponential time,
        $\p^{A}=\bpp^{A}$. We also show that  almost all languages in \pspace\ do not have small  nonuniform complexity.

        We then switch to the  second Baire category notion (called locally-computable),
        and show that the class \sparse\ is meager in \p .
	We show that in contrast to the resource-bounded measure case,
	meager-comeager laws can be obtained for many standard complexity classes,
	relative to locally-computable Baire category on \bpp\ and \pspace .

        Another topic where locally-computable Baire categories differ from resource-bounded measure
	is regarding weak-completeness:
	we show that there is no weak-completeness notion in \p\ based on locally-computable Baire categories,
        i.e. every  $\p$-weakly-complete  set  is complete for \p . We also prove that the  class
        of complete sets for \p\ under Turing-logspace reductions is meager in \p ,
	if \p\ is not equal to $\dspace (\log n)$, and that the same holds unconditionally
	for \qp .

	Finally we  observe that locally-computable Baire categories are incomparable with all existing resource-bounded measure
	notions on small complexity classes, which might explain why those two settings seem to  differ so fundamentally.

\end{abstract}


\section{Introduction}
	Both resource-bounded  Baire categories and resource-bounded measure were  introduced by Lutz
      	in \cite{Lutz:category_measure,Lutz:almost_everyw_high_nonunif_complex}
	for both complexity classes \e\	and \expc .
	They provide a means of investigating the sizes of various subsets of \e\ and \expc , and
	give a notion of small sets, called meager sets.
	Both  resource-bounded measure and resource-bounded Baire category have been successfully used to understand the structure of
	the exponential time classes \e\ and \expc .

	Similarly to resource-bounded measure \cite{Lutz:category_measure}, 
	Lutz's formulation for Baire categories is a general theory which
	holds  for many standard deterministic complexity classes containing $\e$ ranging from $\espace$ to $\rec$.
	Unfortunately Lutz's formulation does not work on feasible complexity classes contained in \e\ like \p ;
	one reason for this is that the characteristic sequence of a language is exponentially larger than the
	strings whose membership bits it is coding for. Thus simply reading such a characteristic sequence is
	above the computational power of \p . Moreover Baire categories are defined via functions (called finite extension strategies)
	extending characteristic sequences of
	languages, i.e. of the form $h(w)=wu$; thus computing   the image of such a function    is also
	above the computational power of \p .
	Whereas some answers to this problem were proposed for the resource-bounded  measure case
	\cite{b:Strauss_measure_on_P_1,b.Mayordomo.measure.pspace,B:Strauss_measure_on_P_3,Moser:dim-in-P},
	the question was still left open in the
	Baire category setting.

	In this paper, we propose two Baire category notions on small complexity classes, like $\p$, \qp , \subexp\ and \pspace ,
	which differ solely  on how the corresponding finite extension strategies are computed. The idea is that instead of computing the 
	whole image of
	some finite extension strategy $h(w)=wu$, we only require the extension $u$ to be computable in polynomial time.

	Ideally, a measure notion in quantitative complexity should satisfy the \emph{three basic properties}:
   \begin{enumerate} 
		\item	Every  singleton set is small. 
		\item Enumerable infinite unions of small sets are small.
		\item	The whole class is not small.
        
	\end{enumerate}
	These basic properties meet the essence of Lebesgue measure and ensure that no class is both large and small.
	We show that both  Baire category notions introduced in this paper  satisfy the three basic properties.
	
	In the classical setting, Baire categories can be alternatively characterized by Banach-Mazur games
	(see \cite{b:oxtoby}), which are infinite two-player
	games, where each player alternatively extends the string output by the other player. 
	We show that Baire category notions on small complexity classes can also be recharacterized in terms of Banach-Mazur games,
	similarly to Baire categories on $\expc$ \cite{Lutz:category_measure}.

	There is another limitation of Lutz's formulation of Baire category  \cite{Lutz:category_measure}
	(which also occurs in resource-bounded measure 
	\cite{b:Regan_probabilistic_martingales,Moser:dim-in-P}): 
	it  works well for deterministic
	classes, but not for probabilistic ones.  We remedy this situation by introducing a  Baire category notion
	on the class \bpp .

	As an application of the first notion, we answer a variant of a question raised in \cite{b:Strauss_measure_on_P_1}, by
	showing that almost all (all except a meager class)
	languages computable in subexponential
	time, are hard enough to derandomize \bpp , i.e. a polynomial time algorithm can use almost
	any language  $L\in\subexp$ to derandomize every probabilistic polynomial time algorithm,
	even if the probabilistic algorithm also has oracle access to $L$
	(whereas in \cite{b:Strauss_measure_on_P_1}, the probabilistic algorithm has no access to $L$).
	We also investigate the nonuniform complexity of languages in \pspace ,
	and  show that almost all languages in \pspace\ do not have small nonuniform complexity.
	A preliminary version of the  first Baire category notion introduced in this paper was published in \cite{b:moser_categories_P}.

	Although the first  Baire category  notion introduced here has interesting applications, the class of languages of
	subexponential density
	is not small relative to it.
	To overcome this, Section \ref{c.localcategory}  introduces a second, stronger, Baire category notion,  called locally-computable.
	The second notion is an adaptation of a previous improvement of Lutz's 
	\cite{Lutz:category_measure,Lutz:almost_everyw_high_nonunif_complex}
	notion by Fenner \cite{Fenner:rb_local_category}, to the setting of small complexity classes.
	The idea in \cite{Fenner:rb_local_category} is to consider finite extension strategies whose image is locally computable in
	a given time-bound instead of globally computable, which yields  a
	stronger Baire category notion on the class \e .
	Similar to Lutz's \cite{Lutz:category_measure,Lutz:almost_everyw_high_nonunif_complex}
	notion, the Baire category notion of \cite{Fenner:rb_local_category}
	only holds on deterministic complexity classes above (and including) \e .
	In Section \ref{c.localcategory} we extend Baire categories from \cite{Fenner:rb_local_category}, (called local categories) to small complexity 	classes like \p , \qp , \subexp , \pspace\ and \bpp .

	Informally speaking, a class is said to be  meager  if there is a 
	computable finite extension strategy
	that given the prefix of the characteristic sequence of any language in the class, extends it to a string which is no longer a prefix of
	the characteristic sequence of the language. In Section \ref{c.category}
        the extension of the finite extension strategy is  required to be polynomial time
        computable.
        For locally computable finite extension strategies, we only require the extension to be bit-wise
        polynomially computable, similarly  to \cite{Fenner:rb_local_category}. 
	This means that the output of locally computable finite extension strategies can be of any finite size,
	which yields  a stronger resource-bounded Baire category notion than the one from  Section
        \ref{c.category}: the class of languages with subexponential density is meager,
        relative to this second Baire category notion.

	Next we investigate  meager-comeager laws in the Baire category setting.
	A well studied topic in resource-bounded measure deals with understanding which subclasses of
	\expc\ satisfy the zero-one law, i.e. classes that have either measure zero or one in \expc .
	Zero-one laws were obtained for all three probabilistic classes  \bpp , \rp\ and \zpp\
	\cite{Melkebeek:zero_one_law,b:moser_0-1_law_for_RP}.
	These laws 
	tell us that  either probabilistic algorithms are in some sense weak, or randomness is intractable.
	Recently a small-or-large law for \spp\ was proved in \cite{b.hitchcock_spp}.
	Although  resource-bounded measure notions were introduced
	on almost all  small complexity classes
	\cite{b:Strauss_measure_on_P_1,b.Mayordomo.measure.pspace,B:Strauss_measure_on_P_3,Moser:dim-in-P},
	no zero-one law has been obtained for the measure notion on \pspace , nor the measure on \bpp\ yet,
	i.e. we have no example of classes which have either measure zero or one in \pspace\ (nor in \bpp ).
	We show that for  local Baire category  on \pspace\ things are different: every standard class contained in
	\pspace\ is either meager in  \pspace\ or equal to \pspace . The same holds for replacing  \pspace\ with \bpp ,
	yielding that  either derandomization is possible i.e. $\bpp = \p$,
	or $\p$ is small compared to $\bpp$.

	\begin{longtable}{|c|c|c|}
		\caption{Classes Satisfying a Small-Large Law}\\
		\hline
		\endhead
		\hline
		\endfoot
		Zero-one Laws  in & for Resource-bounded Measure & for Baire Category\\
		\hline\hline
		\e & \zpp , \rp , \bpp\ \cite{b:moser_0-1_law_for_RP,Melkebeek:zero_one_law} & \zpp , \rp , \bpp , \np\ \cite{Fenner:rb_local_category} \\
		\hline
		\subexp & \zpp , \rp , \bpp\ \cite{Moser:dim-in-P} & \zpp , \rp , \bpp , \np\ $[\mathrm{Section } \ \ref{s.all-meager-pspace}]$ \\
		\hline
		\pspace & ? & \p , \zpp , \rp , \bpp , \np\ $[\mathrm{Section }$ \ref{s.all-meager-pspace}] \\
		\hline
		\bpp & ? & \p , \zpp , \rp \ $[\mathrm{Section }$ \ref{s.all-meager-pspace}]\\
		\hline
	\end{longtable}

	Another area where resource-bounded measure and Baire categories on small complexity classes seem to differ
	is regarding weak-completeness.
        A language $A$ in some class $C$ is said to be $C$-weakly-complete \cite{Lutz:weaklyhard-sets} if
        its lower span, i.e. the class of sets reducible to $A$, does not have $C$-measure zero.
        Lutz showed in \cite{Lutz:weaklyhard-sets} the existence of $\e$-weakly-complete sets  that are not $\e$-complete.
        Similarly we can define a categorical weak completeness notion, by calling a set
        loc-weakly-complete if its lower span is not loc-meager.
	Whereas it is not known whether strictly $\p$-weakly-complete sets (i.e sets that are $\p$-weakly-complete but not $\p$-complete) 
	exist in the resource-bounded measure setting,
	we show that strictly $\p$-loc-weakly-complete languages do not exist, 
        i.e. every $\p$-loc-weakly-complete language
        is also $\p$-complete.

        We also prove that the class of complete languages for \p\ under
        Turing-logspace reductions is loc-meager in \p , if \p\ is not equal to $\dspace (\log n)$,
	and that the same holds unconditionally
	for \qp , contrasting with the lack of any such result in the resource-bounded measure setting.

	Finally we observe that  locally-computable  Baire categories on \p\ are incomparable to
	the resource-bounded measure notions on \p\ from
	\cite{Moser:dim-in-P},
	in the sense that every set which is random for \p -computable martingales
	is meager for local categories on \p ; and that there are generic sets for local categories
	on \p\  which have  \p -measure zero. This shows that the size notion derived from
	both \p -measure and local categories on \p\ differ fundamentally
	(the same holds for \qp , \subexp , \ldots, \e ),
	which might explain why most of our applications are not known to hold
	in the setting of resource-bounded measure on small complexity classes.

\section{Preliminaries}

	For complexity classes we use the notation from
	 \cite{Diaz:structural_complexity1,Diaz:structural_complexity2,Papadimitriou:comp_complexity}.
	To give a general theory on small complexity classes, we  use the following formalism.
	A family of time bounds is a set of functions $\Delta \subset \{ t: \N \rightarrow \N, t \text{ is computable }  \} $. 
	The time bounds we shall consider are 
	$\mathrm{poly}= \cup_{k\in \N} \, O(n^k) $, 
	$\mathrm{quasipoly}= \cup_{k\in \N}\, O( n^{\log^k n} )$,
	$\mathrm{quasipolylin}= \cup_{k\in \N} \, O( n^{k\log n}) $ and
	$\mathrm{subexp_{\epsilon}}= \cup_{\delta <\epsilon} \, O(2^{n^{\delta}})$ (where $0<\epsilon <1$),
	and let 
	$$\Small = \{ \mathrm{poly}, \mathrm{quasipoly},\mathrm{quasipolylin},\mathrm{subexp_{\epsilon}} \}.$$
	For a family of time bounds $\Delta\in \Small$, we define its corresponding (small) complexity
	classes $\Tc(\Delta)= \cup_{t\in\Delta} \dtime (t)$, 
	$\Sc(\Delta)= \cup_{t\in\Delta} \dspace (t)$ and
	$\BPc(\Delta)= \cup_{t\in\Delta} \bptime (t)$. 
	The (small) complexity classes we 
	shall be interested in are 
	$\p = \Tc(\mathrm{poly})$, 
	$\qp = \Tc(\mathrm{quasipoly})$,
	$\qplin = \Tc(\mathrm{quasipolylin})$,
	$\e_{\epsilon} = \Tc(\mathrm{subexp_{\epsilon}})$ (where $0<\epsilon <1$),
	$\subexp=\cap_{\epsilon >0}\e_{\epsilon}$, 
	$\pspace = \Sc(\mathrm{poly})$, and
	$\bpp = \BPc(\mathrm{poly})$.

	Regarding \qplin\ and \qp , notice that whereas it is easy to show that the canonical complete language
	(i.e. whose strings are of the form a padding followed by an index $i$ and a string $x$, such that the $i$th machine
	accepts string $x$ in at most $t$ steps, where $t$ depends on the size of the padding)  is complete for
	\qplin , it is not clear whether it is  complete for \qp .

	Let us fix some notations for  strings and languages.
	A \emph{string} is an element of $\bool{n}$ for some integer $n$.
	For a string $x$, its length is denoted by $|x|$.
	$s_0 ,s_1 , s_2 \ldots $ denotes the standard enumeration of the strings in $\bool{*}$ ordered 
	by length and then  lexicographically, where $s_0 = \lambda$ denotes the empty string. 
	Note that $|w|=2^{O(|s_{|w|}|)}$.
	For a string $s_i$ define its position by $\pos(s_i)=i$.
	If $x,y$ are strings, we write $x\leq y$ if $|x|<|y|$ or $|x|=|y|$ and $x$ precedes $y$ in lexicographical order.
	A \emph{sequence} is an element of $\bool{\infty}$.
	If $w$ is a string or a sequence and $1\leq i \leq |w|$
	then $w[i]$ and $w[s_i]$  denote the $i$th bit of $w$.
	Similarly $w[i \ldots j]$ and $w[s_i \ldots s_j]$ denote the $i$th through $j$th bits. $\dom(w)$ denotes
	the domain of $w$, where $w$ is viewed as a partial function.

	For two string $x,y$, the concatenation of $x$ and $y$ is denoted $xy$.
	If $x$ is a string and $y$ is a string or a sequence extending $x$ i.e. $y=xu$, where $u$ is a string or a sequence,
	we write $x \sqsubseteq y$.
	We write $x \sqsubset y$ if $x \sqsubseteq y$ and $x \neq y$.

	A \emph{language} is a set of strings.
	A \emph{class} is a set of languages.
	The cardinality of a language $L$ is denoted $|L|$.
	Let $n$ be any  integer. The set of strings of size $n$ of language $L$ is denoted $L^{=n}$.
	Similarly $L^{\leq n}$ denotes the set of strings in $L$ of size at most $n$.
	We identify a language $L$ with its characteristic function $\chi_L$, where
	$\chi_L$ is the sequence such that $\chi_L[i] =1$ iff $s_i \in L$.
	Thus a language can be seen as a sequence in $\bool{\infty}$.

	Let $A$ be any language. The lower span (resp. upper span) of $A$,
	denoted $A^{\geq^p_m}$  (resp. $A^{\leq^p_m}$) is the set of languages $B$ such that $B\leq^p_m A$ (resp. $A\leq^p_m B$).

	For $a,b \in \N$ let $a \moins b$ denote $\max (a-b , 0)$.

	\subsection{Pseudorandom Generators}
	The hardness of a generator is the size of the smallest circuit which can distinguish the output
	of the generator from truly random bits. More precisely,
	\begin{definition}
	         Let $A$ be any language. The hardness $H^{A}(G)$
                of a random generator
                $G :
                \{ 0 , 1 \} ^{m}
                \longrightarrow
                \{ 0 , 1 \} ^{n}$, is defined as the minimal $s$ such that
                there exists an $n$-input circuit $C$ with
                oracle gates to A, of size at most s, for which:
                $$ \left| \Pr_{x \in \{ 0, 1 \}^{m}}
                [C(G(x))=1] -
                 \Pr_{y \in \{0, 1\}^{n}}
                [C(y)=1] \right| \geq \frac{1}{s} \mbox{.}$$
        \end{definition}
	The $A$-oracle circuit complexity of a Boolean function $f$,
	denoted $\size^A(f)$ is defined as the size of the smallest circuit with oracle to $A$
	computing $f$.
	It was discovered in \cite{melkebeek:graphnonisomorphism}
	that the construction of pseudorandom generator from high circuit complexity  Boolean
	functions does relativize.
        \begin{theorem}[Klivans-Melkebeek]    \label{t:kliv_mel}
              	Let $A$ be any language. There is a polynomial-time computable function
		$F:\{ 0,1\}^{*} \times \{ 0,1\}^{*} \rightarrow \{ 0,1\}^{*}$,
		with the following properties. For every $\epsilon >0$, there exists
		$a,b \in \mbb{N}$ such that for any $n\in\N$
		$$F:\{ 0,1\}^{n^{a}} \times \{ 0,1\}^{b\log n} \rightarrow \{ 0,1\}^{n} \mbox{,}$$
		and if	 $r$ is the truth table of a $(a \log n)$-variables Boolean function of A-oracle circuit
		complexity at least $n^{\epsilon a}$, then the function
		$G_r(s) = F(r,s)$ is a generator, mapping
		$\{ 0,1\}^{b\log n}$ into $\{ 0,1\}^{n}$, which has hardness
		$H^{\mrm{A}}(G_{r}) >n$.
        \end{theorem}

	\subsection{Finite Extension Strategies}

	Whereas resource-bounded measure is defined via  martingales,  resource-bounded Baire categories require  finite extension
	strategies. Here is a definition.

	\begin{definition}
		A function $h: \bool{*} \rightarrow \bool{*}$ is a finite extension strategy, or a constructor,
		if for every string
		$\tau \in \bool{*}$, $\tau \sqsubseteq h(\tau )$.
	\end{definition}
	For simplicity we  use the word \emph{strategy}  for finite extension strategy.
	We say a strategy $h$ avoids some language $A$ (or language $A$ avoids strategy $h$)
	if for every  string $\tau \in \bool{*}$ we have
	$$h(\tau ) \not \sqsubseteq \chi_{A} \, . $$
	We say a strategy $h$ meets some language $A$
	if $h$ does not avoid $A$.

	We  often consider indexed strategies. An indexed strategy is a function
	$$h: \N \times \bool{*} \rightarrow \bool{*}$$
	such that $h_i := h(i,\cdot )$ is a strategy for every $i\in \N$.
	Let $h$ be an indexed strategy. Consider the following function $\ext$.
	Let $\sigma \in \bool{*}$ and $i,k\in \N$ and let $w$ be the unique string such that
	$h_i (\sigma)= \sigma w$. Define
	$$\ext (h_i(\sigma),k) = \left\{
					\begin{array}{ll}
						w[k] & \mbox{if } 1\leq k \leq |w| \\
						\perp & \mbox{otherwise.}
						\end{array}
		            	  \right.
	$$
	and
	$$\ext (h_i(\sigma)) = w \ .$$


\section{Baire Category on Small Complexity Classes}\label{c.category}
	
	For the rest of this paper, let $\Delta\in\Small$ be a family of bounds and let $\cla= \Tc (\Delta)$. 
	be the corresponding small time complexity class (for instance $\cla = \p$).
	Note that most results in this paper
	also hold for small space-bounded classes, i.e. of the form $\cla_S= \Sc(\Delta)$ (for instance $\cla_S = \pspace$).

	To define  resource bounded Baire categories on  $\cla$, 
	we  consider strategies computed by
	Turing machines which  have random access to their inputs, i.e. on input $\tau$,
	the machine can query any bit of $\tau$ to its oracle.
	For such a  random access Turing machine $M$ running on input $\tau$, we denote this convention by
	$M^{\tau}(\cdot)$.
	Note that random access Turing machines can compute the lengths of their input $\tau$ in $O(\log |\tau|)$ steps
	(by checking $\tau[1],\tau[2], \tau[2^2],\cdots ,\tau[2^i]$ until they go outside the string,
	followed by a binary search). We  consider random access Turing machines running in time
	$t(|s_{|\tau |}|)$ (equivalently $t( \log |\tau|)$ for some $t\in\Delta$. 
	Nevertheless, for the time-bounded case, such  machines cannot
	read their entire input (because time bounds in $\Delta$ are less than exponential).

	The idea to define a Baire category notion on $\cla$, is to consider strategies whose extension is
	computable in time  $t\in\Delta$, instead of requiring the whole output to be computable, which would
	not be possible in time $t\in\Delta$.
	\begin{definition}\label{d:pcompindexedstratgy}
		An indexed  strategy $h: \N \times \bool{*} \rightarrow \bool{*}$ is $\Delta$-computable 
		if there is a random access Turing machine $M$ as before, such that for every $\tau \in \bool{*}$
		and every $i \in \N$,
		\begin{equation}
			M^{\tau}(i)=\ext( h_i (\tau))
		\end{equation}
	where $M$ runs in time  $t(\log |\tau| + |i|)$, for some $t\in\Delta$. For the space bounded case, we also require the output tape
	to be bounded, i.e.
	$|\ext( h_i (\tau))| \leq t(\log |\tau| + |i|)$.
	\end{definition}

	We say a class is small if there is an  indexed strategy that avoids every language in the class.
	More precisely,

	\begin{definition}
		A class $X$ of languages is $\cla$-meager  if there exists a $\Delta$-computable 
		indexed
		strategy $h$, such that for every $L \in X$ there exists an index  $i$ such that
		$h_i$ avoids $L$.
	\end{definition}

	A class is called comeager if its complement is meager.

	In order to formalize the second basic property we need to define \emph{enumerable infinite unions} precisely.

	\begin{definition} \label{d:finiteunions}
		$X=\bigcup_{i\in\N}X_i$ is a $\cla$-union  of $\cla$-meager sets, 
		if there exists an indexed
		$\Delta$-computable  strategy
		$$h:\N \times \N \times \bool{*} \rightarrow \bool{*}$$ 
		(i.e. for any $i,j\in\N, \tau\in\bool{*}$, $h(\tau, i,j)$ is computable in time  $t(\log |\tau| + |i|+|j|)$, for some $t\in\Delta$)
		such that for every
		$i \in \N$, $h_{i,\cdot}$ witnesses $X_i$'s meagerness.
	\end{definition}


	Let us prove the three basic properties.

	\begin{theorem} \label{t:ax1p}
		For any language L  in $\cla$,  the singleton $\{ L\} $ is $\cla$-meager.
	\end{theorem}
	\begin{proof}
		Let $L \in \cla$  be any language. We describe  a $\Delta$-computable  constructor $h$ which
	avoids $\{ L\}$. Consider the following Turing machine $M$ computing $h$.
	On input string $\sigma$,
	$M^{\sigma}$ simply outputs $1-L(s_{|\sigma |+1})$.
	$h$ is clearly  $\Delta$-computable, and $h$ avoids $\{ L\}$.
	\qed
	\end{proof}

	The proof of the second basic property is easy.
	\begin{theorem} \label{t:ax2p}
		A $\cla$-union of $\cla$-meager  sets is $\cla$-meager.
	\end{theorem}
	\begin{proof}
		It is easy to see that a $\Delta$-computable  strategy
		$$h:\N \times \N \times \bool{*} \rightarrow \bool{*}$$ can be transformed into a 
		$\Delta$-computable  strategy
		$$h':\N \times \bool{*} \rightarrow \bool{*}$$ using a pairing function to combine
		$\N\times\N$ into $\N$.
	\qed
	\end{proof}

	Let us prove the third basic property which says that the whole space $\cla$  is not small.
	The idea of the proof is given a strategy $h$, construct a language $L$ that meets it, where
	$L$'s characteristic sequences is divided into blocks of exponential size, where block $i$ is used to meet $h_i$.
	\begin{theorem} \label{t:ax3p}

		$\cla$ is not $\cla$-meager.

	\end{theorem}
	\begin{proof}
	Let $h$ be an indexed $\Delta$-computable constructor   
	and let $M$ be a Turing machine computing $h$,
	running in time $t\in \Delta$.
	We construct a language $L \in \cla$  which meets $h_i$ for every $i$.
	The idea is to construct a language $L$ with the following characteristic function,

	\begin{equation}
		\chi_L = \und{0}_{B_0}\und{ \ext( h_1(B_0))0 \cdots 0}_{B_1}
		 \cdots
		\und{\ext (h_i(\boite{i-1}))0\cdots 0}_{B_i}
	\end{equation}
	where block $B_i $ corresponds to all strings of size $i$, and
	block $B_i$ contains
	$$\ext (h_i(\boite{i-1}))$$ followed by a padding with 0's.
	$B_i$ is large enough to contain
	$\ext (h_i(\boite{i-1}))$ because $M$ runs in time $t$, therefore
	$|\ext (h_i(\boite{i-1}))|\leq t(\log (|\boite{i-1}|)) < 2^i$ because $t\in\Delta\in\Small$
	(for the space-bounded case, it follows from the requirement on the size of the output tape, 
	see Definition \ref{d:pcompindexedstratgy}).

	Let us construct Turing machine $N$ deciding $L$, running in time $t'\in\Delta$. 
	On input $x$, where $n=|x|$,
	\begin{enumerate}

		\item Compute $p$ where $x$ is the $p$th word of length $n$.

		\item For $i=1$ to $n$  simulate $M^{\boite{i-1}}(i)$. Answer $M$'s queries with
			the previously stored binary sequences $\bar{B}_1, \bar{B}_2 , \bar{B}_{i-1}$ in the
			following way. Suppose that during its simulation $M^{\boite{i-1}}(i)$ queries the $k$th bit
			of $\boite{i-1}$ to its oracle.
			To answer this query, compute the position $p_k$ of $s_k$ among words of size
			$|s_k|$.
			Look up whether the stored binary sequence $\bar{B}_{l_k}$ contains a $p_k$th bit $b_k$.
			If this is the case answer $M$'s query with $b_k$, else answer $M$'s query with 0.
			Finally store the output of $M^{\boite{i-1}}(i)$ under $\bar{B}_i$.

		\item If the stored binary sequence $\bar{B}_n$ contains a $p$th bit then output this bit,
			else output 0 ($x$ is in the padded zone of $B_n$).

	\end{enumerate}

	Let us check that $L$ is in $\cla$.
	The first and third step are clearly computable in time  $O(t(n))$.
	For the second step we have that for each of the $n$ recursive steps there are at most
	$t(n)$ queries and each simulation of $M$
	once the queries are answered takes
	at most time $t(n)$. Thus $L$ is computable in time  $t^4$, 
	i.e. $L\in\cla$. Note that all $\bar{B}_i$'s have size at most $t(n)$,
	therefore it's no problem to store them.
	\qed
	\end{proof}

	\begin{remark}
		Notice that  as opposed to existing notion of resource-bounded measure on small classes
		\cite{b:Strauss_measure_on_P_1,B:Strauss_measure_on_P_3,Moser:dim-in-P},
		this Baire category notion does not need dependency sets, i.e. polynomial printable sets  corresponding to
		the bits read in the input (whereas this restriction is needed for locally-computable categories).
	\end{remark}


\subsection{Resource-Bounded Banach-Mazur Games} \label{s.bm_p}

	We give an alternative characterization of small sets via resource-bounded Banach-Mazur games,
	similarly to the classical case (see \cite{b:oxtoby}) and the resource-bounded case \cite{Lutz:category_measure}.
	Informally speaking, a Banach-Mazur game, is a game between two strategies $f$ and $g$,
	where the game begins with the empty string $\lambda$.
	Then  $g\circ f$ is applied successively on $\lambda$.
	Such a game yields a unique infinite string, or equivalently a language,
	called the result of the play between $f$ and $g$.
	We say that $g$ is a winning strategy for class $X$ if it forces the result of the game
	with any strategy $f$ to be a language not in $X$.
	We show that the existence of a winning strategy for $X$ is equivalent to the meagerness of $X$.
	This equivalence result is useful in practice, since it is often easier to find a winning strategy
	rather than a finite extension strategy.

	\begin{definition} \text{   }
		\begin{enumerate}
			\item	A play of a Banach-Mazur game is a pair $(f,g)$ of strategies such that
					$g$ strictly extends every string, i.e.
					for every string $\tau \in \bool{*}$, $\tau \sqsubset g(\tau)$.
			\item	The result $R(f,g)$ of the play $(f,g)$  is the unique element of
				$\bool{\infty}$ that extends $(g\circ f)^{i}(\lambda)$ for every $i \in \N$.
		\end{enumerate}
	\end{definition}

	For a class of languages $X$ and time bound families $\Delta_I$ and $\Delta_{II}$,
	denote by $G[X, \Delta_I,\Delta_{II}]$  the Banach-Mazur game with distinguished
	set $X$, where player $i$ ($i\in\{I,II\}$) must choose a strategy $\Delta_i$-computable. 
	We say player II wins the play $(f,g)$ if $R(f,g) \not \in X$, otherwise we
	say player I wins. We say player II has a winning strategy for the game $G[X, \Delta_I,\Delta_{II}]$,
	if there exists a $\Delta_{II}$-computable   
	strategy $g$  such that for every $\Delta_{I}$-computable  strategy $f$,
	player II wins $(f,g)$. We denote by $\N^{\N}$ the class of all functions mapping strings to strings.

	The following result states that a class is $\cla$-meager iff
	there is a winning strategy for player II. The idea of the proof is to construct a non-indexed strategy
	from an indexed one where at step $i$, the smallest index that has not been met is used. 
	\begin{theorem} \label{t:bmgp}
		Let $X$ be any class of languages. The following are equivalent.
		\begin{enumerate}
			\item 	Player II has a winning strategy for $G[X,\N^{\N} , \Delta]$.
			\item	$X$ is $\cla$-meager.
		\end{enumerate}
	\end{theorem}
	\begin{proof}
	Suppose the first statement holds and let $g$ be a  $\Delta$-computable 
	wining strategy for player II.
	Let $M$ be a Turing machine computing $g$, in time $t\in\Delta$.
	We define an indexed $\Delta$-computable  constructor $h$. For $k\in \N$ and $\sigma \in \bool{*}$, define
	\begin{equation}
		h_k(\sigma):= g(\sigma ') \quad \mbox{where } \sigma ' = \sigma  0^{k \moins |\sigma |} \mbox{ .}
	\end{equation}

	$h$ is $\Delta$-computable    because computing $h_k(\sigma)$ simply requires simulating
	$M^{\sigma '}$, answering M's   queries in  $\dom(\sigma ')\backslash \dom(\sigma)$
	by $0$.
	We show that if language $A$ meets $h_k$ for every $k \in \N$, then $A \not \in X$.
	This implies that $X$ is $\cla$-meager as witnessed by $h$. To do this we show that
	for every $\alpha \sqsubset \chi_A$ there is a string $\beta$ such that,
	\begin{equation}
		\alpha \sqsubseteq \beta \sqsubseteq g(\beta) \sqsubset \chi_A \mbox{ .}
	\end{equation}
	If this holds, then player I has a winning strategy yielding $R(f,g) = A$ (unless $A\not\in X$):
	for a given $\alpha$ player I extends it to obtain the corresponding $\beta$, thus forcing
	player II to extend to a prefix of $\chi_A$.
	So let $\alpha$ be any prefix of $\chi_A$, where $|\alpha| =k$.
	Since $A$ meets $h_k$, there is a string $\sigma \sqsubset \chi_A$   such that
	\begin{equation}
		 \sigma ' \sqsubseteq g(\sigma ') = h_k(\sigma) \sqsubset \chi_A
	\end{equation}
	where $\sigma ' = \sigma  0^{k\moins |\sigma| }$.
	Since $|\alpha | \leq |\sigma '|$ and $\alpha$, $\sigma '$ are prefixes of $\chi_A$, we have
	$\alpha \sqsubseteq \sigma '$.
	Define $\beta$ to be $\sigma '$.

	For the other direction, let $X$ be $\cla$-meager as witnessed by $h$, i.e.
	for every $A \in X$ there exists $i \in \N$ such that $h_i$ avoids $A$. Let $N$ be a Turing machine
	computing $h$, running in time  $t\in\Delta$.
	We define a $\Delta$-computable   constructor $g$  inducing a winning strategy for player II in the game
	$G[X,\N^{\N} , \Delta]$. 
	We show that for any strategy $f$, $R(f,g)$ meets $h_i$ for every
	$i\in \N$, which implies  $R(f,g) \not \in X$.
	Here is a description of a Turing machine $M$ computing $g$.
	For a string $\sigma$, $M^{\sigma}$ does the following.
	\begin{enumerate}
		\item 	\label{i:una}Compute $n_0 = 
			\min \{ t: t \leq  \log |\sigma|, \text{ and } (\forall \tau \sqsubseteq \sigma \mbox{ such that }
			|\tau |\leq \log |\sigma|) \ \ h_t (\tau) \not \sqsubseteq \sigma \}$.
		\item	If no such $n_0$ exists output $0$.
		\item	If $n_0$ exists ($h_{n_0}$ is the next strategy to be met),
			simulate $N^{\sigma }(n_0)$, denote $N$'s answer by $\omega$.
			Output $  \omega$.
	\end{enumerate}

	$g$ is computable in  time $O(n^2 t(n))$, i.e.  $\Delta$-computable.
	We show that $R(f,g)$ meets every $h_i$ for any strategy $f$. Suppose for a contradiction that
	this is not the case, i.e.
	there is a strategy  $f$ such that $R(f,g)$ does not meet $h$.
	Let $m$ be the smallest index such that $R(f,g)$ does not meet $h_{m}$.
	Since $R(f,g)$ meets $h_{m -1}$ there is a string $\tau$ such that
	$$h_{m -1}(\tau) \sqsubset R(f,g) \ .$$
	Since $g$ strictly extends strings at every round, after at most $2^{O(|\tau |)}$ rounds,
	$f$  outputs a string $\sigma$ long enough to enable step one (of $M$'s description) to
	find out that
	$$h_{m -1}(\tau) \sqsubseteq \sigma \sqsubset R(f,g) $$
	thus incrementing $m -1$ to $m$.
	At this round we have
	$$g(\sigma)= \sigma  \, \ext (h_{m}(\sigma  ))$$
	i.e.
	$$h_{m} \sqsubset R(f,g)$$ which is a contradiction.
	\qed
	\end{proof}


\subsection{Application: Derandomization of BPP} \label{s.application}

        It was shown in \cite{b:Strauss_measure_on_P_1} that for every $\epsilon >0$,
        for all  languages $A \in \e_{\epsilon}$ except a measure zero class
        $\bpp$ is contained in
        $\p^A$, thus leaving open the question whether the result also
		  holds if the probabilistic algorithms also have access to $A$.
		  We answer this question affirmatively in the Baire category setting, i.e. we show that for every
        $\epsilon >0$, all languages
        $A\in \e_{\epsilon}$ except a meager class
        satisfy
		  $\p^A =\bpp^A$.
	The idea of the proof is to construct a strategy that given a initial segment of a language, extends it by 
	the characteristic sequence of a language with high circuit complexity, by diagonalizing over all small circuits.
	The language can then be plugged  into a pseudorandom generator to obtain full derandomization.
	\begin{theorem}

		For every $\epsilon>0$, the set of languages $A$ such that
		$ \p^{A}\neq\bpp^{A}$ is $\e_{\epsilon}$ -meager.

	\end{theorem}
	\begin{proof}
	Let $\epsilon > 0$ be small.
	Let $0<\delta <\epsilon$ and let $b>0$ be some integer to be determined later. 
	Consider the following strategy $h$, computed by the following Turing machine $M$.
	On input $\sigma$, where $n=\log |\sigma |$, $M$ does the following.
	At start $Z=\emptyset$, and $i=1$. $M$ computes $z_i$
	in the following way. First if 
	$$s_{|\sigma| +i} \neq 0^{2^{b|u|}}u$$
	for  any string $u$ where
	$|u|=\log (n^{1/b})$,  then let $z_i = 0$, output $z_{i}$, and compute $z_{i+1}$;
	else denote by $u_i$ the corresponding string $u$. Construct  the set
 	$T_i$ of all truth tables of $|u_i|$-input Boolean circuits $C$ with oracle gates for $\sigma$
	of size less than $2^{\delta |u_i|}$, such that
	$$C(u_j)=z_j \ \text{ for every } \ (u_j ,z_j)\in Z \ .$$
	Compute $$M_i = \text{Majority}_{C\in T_i}[C(u_i)]$$ and let $z_i = 1-M_i$. Add $(u_i , z_i)$ to $Z$.
	Output $z_i$, and compute $z_{i+1}$, unless $u_i = 1^{\log (n^{1/b})}$ (i.e. $u_i$ is the last string
	of size $\log (n^{1/b})$), in which case $M$ stops.

	There are $2^{n^{4\delta /b}}$ circuits to simulate,
	and simulating such a circuit takes time $O(n^{4\delta/b})$, by answering its queries to $\sigma$
	with the input $\sigma$.
	Finally computing the majority $M_i$  takes time $2^{O(n^{4\delta /b})}$.
	Thus the total running time is less than $2^{n^{2c\delta / b}}$ for some constant $c$,
	which is less than $2^{n^{\epsilon '}}$ (with $\epsilon ' < \epsilon $)
	for an appropriate choice of $b$.

	Let $A$ be any language and consider
	$$F(A):= \{ u|0^{2^{b|u|}}u \in A \} .$$
	It is clear that $F(A) \in \e^{A}$.
	Consider $H^A_{\delta}$ the set of languages $L$ such that every
	$n$-input circuits with oracle gates for $A$ of size less than $2^{\delta n}$ fails to compute $L$.
	We have
	$$F(A) \in H^{A}_{\delta} \ \text{ implies } \  \p^{A} = \bpp^{A}$$
	by Theorem \ref{t:kliv_mel}.

	We show that $h$ avoids every language $A$ such that
	$$F(A)\not\in H^{A}_{\delta} \, .$$
	So let $A$ be any such language.  There is a
	$n$-inputs circuit family
	$\{C_n \}_{n>0}$, with oracle gates for $A$, of size less than $2^{\delta n}$ computing $F(A)$.
	We have
	\begin{equation}\label{e:2}
		C(u_i)=1 \text{ iff } 0^{2^{b|u_i|}}u_i \in A \ \text{ for every string } u_i
		\text{ such that } (u_i,z_i) \in Z.
	\end{equation}
	(for simplicity we omit
	$C$'s index). Consider the set
	$D_n$ of all circuits with $\log (n^{1/b})$-inputs of size at most
	$n^{\delta /b}$ with oracles gates for $A$ satisfying Equation \ref{e:2}.
	We have $|D_n| \leq 2^{n^{4\delta /b}}$.
	By construction,  every $z_i$ such that $(u_i, z_i) \in Z$ reduces the cardinal of $D_n$
	by a factor $2$. Since there are $n^{1/b}$ $z_i$'s such that $(u_i,z_i)\in Z$,
	we have
	$$D_n \leq 2^{n^{4\delta /b}} \cdot 2^{-n^{1/b}} $$
	which is less than one because $\delta$ is small (smaller than $1/4$).
	i.e. $D_n = \emptyset$.
	Therefore $h(\sigma) \not\sqsubset \chi_{A}$.
	\qed
	\end{proof}


\subsection{Almost every Language in PSPACE Does Not Have Small Nonuniform Complexity} \label{s.pspacelargecircuits}

	The following result shows that almost every language in \pspace\ does not have small nonuniform complexity.
	\begin{theorem}
		Let  For every $c>0$, $\size (n^c)$ is $\pspace$-meager.
	\end{theorem}

	\begin{proof}
		Let $c>0$. Consider the following strategy which on input $\sigma$ with $n=\log |\sigma|$,
		outputs a string $u$ of size $2n^{c+1}$ defined as follows.
		Let $u_i$ denote the $i$th bit of $u$, and let $z_{u_i}$ be the string whose membership bit
		corresponds to $u_i$, in the characteristic sequence starting with $\sigma u$.
		For  $n\leq t$ denote by $S(n,t, \sigma , u_1 \cdots u_l)$ the set of $n$-inputs  Boolean circuits of size $t$ such that
		$C(z_{u_j})=u_j$ for every $1\leq j \leq l$. Let
		$$u_i= 1- \mathrm{Majority} \{ C(z_{u_i}): C\in S(|z_{u_i}|,|z_{u_i}|^c, \sigma , u_1 \cdots u_{i-1}) \}.$$
		It is well known \cite{Papadimitriou:comp_complexity} that
		$$|S(n,t, \sigma , u_1 \cdots u_t)|\leq 2^{t\log t}$$ thus
		$h$ is computable in $\dspace (n^{c+2})$ by constructing all corresponding circuits.
		$h$ avoids $\size (n^c)$, because there are less than $2^{n^{c+1}}$ such circuits,  and each bit $u_i$ of
		the extension $u$ diagonalizes against at least half such circuits.
		\qed
	\end{proof}

        The proof of Theorem \ref{t:ax3p} shows that the class  of all languages with subexponential density is
        not $\cla$-meager. In the next section, we  improve the power of $\Delta$-computable   
			strategies
        by considering locally computable strategies,
		  which can  avoid
		  the class of languages of subexponential density.


	\section{Locally Computable  Categories on Small Complexity Classes} \label{c.localcategory}

	 In order to allow random access  Turing machines  to compute
	the length of their inputs $\tau$,
	without querying their oracles (due to the query set requirement that follows), 
	we also provide  them with $s_{|\tau|}$.
	For such a Turing machine $M$ running on input $\tau$, we denote this convention by
	$M^{\tau}(s_{|\tau|})$.

        Similarly to \cite{Fenner:rb_local_category}, 
	we shall consider strategies whose extensions are bit-wise computable in time  $t\in\Delta$.
        Such strategies are very strong since the extension can be of any finite size,
        as long as it is locally computable. As it is usually the case with most notions of measure defined in
	small complexity classes \cite{b:Strauss_measure_on_P_1,B:Strauss_measure_on_P_3,Moser:dim-in-P}, 
 	to be able to show that the whole class is not small,
        the power of the Turing machines computing the  strategies  needs to be reduced, by requiring that all queries made to the input
        are contained in a $t$-printable set ($t\in\Delta$), called the query set
		   (a set $S$ is $t$-printable if there is
			a  Turing machine $M$ which on input $1^n$ outputs all strings in $S^{=n}$ in time $t(n)$).

	\begin{definition}
		An indexed   strategy $h: \N \times  \bool{*} \rightarrow \bool{*}$ is called $\Delta$-loc-computable
		if  there exists a random access  Turing machine $M$ as before  such that for every $\tau \in \bool{*}$
		and every $i,k \in \N$,
		$$M^{\tau}(s_{|\tau|}, i,k)=\ext (h_i (\tau),k)$$
		where $M$ runs in time  $t(\log |\tau| + |i| +|k|)$ for some $t\in\Delta$,
		and there is a $t$-printable query set $G$ such that for every
		$n,i,k \in \N$ and for every $i',k' \in \N$ such that $i'\leq i$ and
		$k'\leq k$ and for every input $\sigma \in \bool{*}$ such that  $\log|\sigma| \leq n$,
		$M^{\sigma}(s_{|\sigma|},i',k')$ queries $\sigma$ only on bits that
		are in $G(n,i,k)$, where $G(n,i,k)$ is printable in time 
		$t(n+|i|+|k|)$.
	\end{definition}

        A class of languages is called meager if there is an indexed  strategy that avoids every language in the class.

	\begin{definition}
		A class $X$ of languages is $\cla$-loc-meager  if there exists a $\Delta$-loc-computable
		 indexed
		strategy $h$, such that for every $L \in X$ there exists $i\in \N$, such that
		$h_i$ avoids $L$.
	\end{definition}

        The definition of \emph{enumerable infinite unions} is similar to Definition \ref{d:finiteunions}.

\subsection{The Three Basic Properties}

        Let us check that all three basic properties hold for  locally-computable Baire categories.
	\begin{theorem} \label{t:ax1ploc}
		For any language L  in $\cla$,  the singleton $\{ L\} $ is $\cla$-loc-meager.
	\end{theorem}
	\begin{proof}
	Follows from Theorem \ref{t:ax1p}.
	\qed
	\end{proof}

        The following result states that enumerable infinite unions of small sets are small.
	\begin{theorem}
			A $\cla$  loc-union  of $\cla$  loc-meager sets is $\cla$  loc-meager.
	\end{theorem}
	\begin{proof}
        Similar to Theorem \ref{t:ax2p}.
	\qed
	\end{proof}

         Let us prove the third basic property. The proof idea is similar to that of Theorem \ref{t:ax3p}, i.e. given a strategy $h$
	we construct a language $L$ that meets $h$, where $L$'s characteristic sequence is divided into blocks where on the $i$th
	block, $h_i$ is met. The difference is that now we do not know the size of the extension computed by $h_i$, therefore we 
	first show the existence of a function that bounds the size of $h_i$, which we use to determine the size of block $i$.
	\begin{theorem}\label{t:ax3ploc}

		$\cla$  is not $\cla$  loc-meager.

	\end{theorem}
	\begin{proof}
	We  need the following technical Lemma, that bounds the size of the extensions efficiently.


	\begin{lemma} \label{l:lt1}

		Let $h$ be a $\Delta$-loc-computable  indexed strategy. Then there exists a function
		$f:\N \rightarrow \N$ such that,
		\begin{enumerate}

			\item	For every $\sigma \in \bool{*}$ such that
				$|\sigma | \leq \sum_{j=0}^{i-1}f(j) $ we have $|\ext (h_i (\sigma))| \leq f(i)$,

			\item	$f(0)=1$ and $f(i)\geq 2^i$ for every $i \in \N$,

		\end{enumerate}

	and there exists a deterministic  Turing machine which on input $i$,
	computes $f(i)$ within $O(\log (f(i)))$ steps.

	\end{lemma}
	\begin{proof}
	Let $h$  be any $\Delta$-loc-computable  indexed strategy and  let $N$ be a Turing machine
	witnessing this fact. We construct a deterministic Turing machine $M$ for $f$.
	At each   step of the computation, $M$ increments a counter $R$.
	On input $i$, $M$ computes $f$ recursively as follows:
	$f(0)=1$. For $i>0$  compute
	$$B= \sum_{j=0}^{i-1}f(j)\ .$$
	For every string $\sigma$ of size at most $B$, simulate
	$N^{\sigma}(s_{\sigma },i,k)$ for $k=1,2, \ldots$ until $N$ outputs $\perp$,
	store  the corresponding $k$ under $k_{\sigma}$.  Compute
	$$K:= \max_{|\sigma| \leq B}k_{\sigma}.$$ Stop incrementing the counter $R$, compute
	$2^{R+K+i}$, and output this value.

	For the running time of $M$ on input $i$, observe that the last two steps (once the counter $R$ is stopped)
	take time $O(R+K+i)$. Thus the total running time of $M$ is at most
	$O(R+K+i)$ which is less than $O(\log (f(i)))$.
	This ends the proof of the Lemma.

        Let us prove the Theorem.
	Let $h$ be a $\Delta$-loc-computable  indexed constructor  and let 
	$M$ be a Turing machine computing $h$
	with query set $G_M$. Let $f$ be as in Lemma \ref{l:lt1} and let $Q$ be a Turing machine
	computing $f$.

	We construct a language $L \in \p$ which meets $h_i$ for every $i$.
	The idea is to construct a language $L$ with the following characteristic function,

	$$\chi_L = \und{0}_{B_0}\und{ \ext (h_1(B_0))0 \cdots 0}_{B_1}
	\und{ \ext (h_2(B_0  B_1 ))0 \cdots 0}_{B_2} \cdots
	\und{\ext (h_i(\boite{i-1}))0\cdots 0}_{B_i}$$
	where block $B_i $ has size $f(i)$ and contains
	$\ext (h_i(\boite{i-1}))$ followed by a padding with 0's.
	$B_i$ is large enough to contain $\ext (h_i(\boite{i-1}))$ by definition of $f$.

	Let us construct a Turing machine $N$ deciding $L$. On input $x$, where $n=|x|$,
	\begin{enumerate}

		\item Compute $\pos(x)$.

		\item	Compute   the index $i(x)$, where the membership bit of $x$ is in
			zone $B_{i(x)}$, with the formula
			$$i(x)=\max_{j\geq 0}[\sum_{t=0}^{j-1}f(t)<\pos(x)+1]$$
			in the following way.
			At the beginning $S=1$, then for $t=1,2,\ldots$ compute $f(t)$ by simulating
			$Q$ for  $|x|^2$ steps, and add the result to $S$, until either $Q$ doesn't
			halt, or $S \geq \pos(x)+1$. Let $t_0$ denote the first $t$ for which this happens,
			and let $$i(x)= t_0 -1 \ .$$

		\item	Compute the position of $x$ in $B_{i(x)}$ with
		        $$\mathrm{rpos}(x)= \pos(x) - F(i(x)-1)$$
			where
			$$F(j) = \sum_{t=0}^{j}f(t) \ .$$

		\item	\label{i:stepprob}Compute the membership bit of $x$,
		        where
			$$\mathrm{bit}(x)=\ext(h_{i(x)}(\boite{i(x)-1}),\mathrm{rpos}(x)) \ .$$
			If  $\mathrm{bit}(x)=\perp$, then output $0$
			($x$ is in the padded zone of $B_{i(x)}$), otherwise output $\bit(x)$.

	\end{enumerate}

	Let us check that $L$ is in $\cla$.
	The first step is clearly computable in time polynomial in $n$.
	For the second step notice that if $f(t)<\pos(x)$, $Q$ must halt within $O(\log (\pos(x)))$ steps,
	which is less than $|x|^2$ since $\pos(x)= 2^{O(|x|)}$. Thus the second step computes $i(x)$
	correctly. Moreover since $f$ increases at least exponentially,  only a polynomial number of terms
	need to be summed in the second and third step.
	Since $f$ is at least exponentially increasing, the sums in step two and three can be done in polynomial time.
	Finally the last step requires simulating
	$$M^{\boite{i(x)-1}}(s_{F(i(x)-1)},i(x), \mathrm{rpos}(x)) \ .$$
	By the hypothesis on $h$,
	$M$'s  queries  are all in
	$$G_M(|s_{F(i(x)-1)}|,i(x), \rpos(x))$$
	which is contained in
	$$G_M(|s_{F(i(x)-1)}|,i(x), \pos(x))$$
	which has size
	 $t(|x|)$ for some $t\in\Delta$. For such a query $q$, i.e. suppose $M$ queries the
	$q$th bit of its input, simply run step one to four above with $x$ replaced by $q$.
	By  definition of $G_M$  at most $t(|x|)$ recursive steps need to be performed.
	\qed
	\end{proof}
	\end{proof}


\subsection{Resource-bounded Banach-Mazur Games} \label{s.bm_ploc}

	Similarly to  Section \ref{s.bm_p} we give an alternative characterization of small sets via resource-bounded Banach-Mazur games.
	The proof is an extension of a similar proof in \cite{Fenner:rb_local_category}.

	\begin{theorem}\label{t:bmploc}
		Let $X$ be any class of languages. The following are equivalent.
		\begin{enumerate}
			\item 	Player II has a winning strategy for $G[X,\N^{\N} , \Delta\text{-}\loc ]$.
			\item	$X$ is $\cla$ loc-meager.
		\end{enumerate}
	\end{theorem}
	\begin{proof}
	We need the following technical Lemma, that gives an efficient bound on the size of the extensions.

	\begin{lemma} \label{l:lt2}

		Let $h$ be a  $\Delta$-loc-computable  indexed constructor. 
		Then there exists a function
		$f:\N \rightarrow \N$ such that for every $m\in \N$, $t\leq m$  and for every string $\tau$
		of size at most $m$,
                $|h_t(\tau)| \leq f(m)$,
		and  there exists a deterministic Turing machine which on input $m$,
		computes $f(m)$ within $O(f(m))$ steps.

	\end{lemma}
	Let us prove the lemma.
        Let $h$ be a $\Delta$-loc-computable  strategy, 
	and let $N$ be a Turing machine witnessing this fact.
	We construct a deterministic Turing machine computing $f$.
	At each   step of the computation, $M$ increments a counter $R$.
	On input $m \in \N$, compute
	$$K = \max_{|\tau |\leq m, t\leq m} \{ |h_t (\tau )|\} $$
	by simulating $N$ on all appropriate strings.
	Stop incrementing the counter $R$, compute $K + R$ and output the result.
	Thus $M$'s total running time is less than $O(R + K)$ which is in
	$O(f(m))$.
        This ends the proof of the Lemma.
	\qed

        For the proof of the Theorem,
	suppose the first statement holds and let $g$ be a wining $\Delta$-loc-computable 
	strategy for player II.
	Let $M$ be a Turing machine computing $g$.
	We define an indexed $\Delta$-loc-computable  
	constructor $h$ by constructing a machine $N$ for $h$;
	let $i\in \N$ and $\sigma \in \bool{*}$,
	$$h_i(\sigma):= g(\sigma ') \quad \mbox{where } \sigma ' = \sigma  0^{i \moins |\sigma |}.$$
	$h$ is $\Delta$-loc-computable because computing $h_i(\sigma)$ simply requires to simulate
	$M^{\sigma '}(s_{|\sigma '|})$ answering M's queries in $\dom(\sigma ' )\backslash \dom(\sigma )$ by $0$.
	Thus $N$'s query set satisfies
	$$G_N(|s_{|\sigma |}| , i ,k) \subseteq G_M(|s_{|\sigma '|}|,k)$$
	which has size 
	$t(\log |\sigma | + |i| + |k|)$ for some $t\in\Delta$, because $|\sigma ' |\leq |\sigma | +i$.

	We show that if language $A$ meets $h_k$ for every $k \in \N$, then $A \not \in X$.
	This implies that $X$ is $\cla$  loc-meager as witnessed by $h$. To do this we show that
	for every $\alpha \sqsubset \chi_A$ there is a string $\beta$ such that,
	$$\alpha \sqsubseteq \beta \sqsubseteq g(\beta) \sqsubset \chi_A.$$
	If this holds, then player I has a winning strategy yielding $R(f,g) = A$:
	for a given $\alpha$ player I extends it to obtain the corresponding $\beta$, thus forcing
	player II to extend to a prefix of $\chi_A$.
	So let $\alpha$ be any prefix of $\chi_A$, where $k=|\alpha| $.
	Since $A$ meets $h_k$, there is a string $\sigma \sqsubset \chi_A$   such that
	$$ \sigma ' \sqsubseteq g(\sigma ') = h_k(\sigma) \sqsubset \chi_A$$
	where $\sigma ' = \sigma  k\moins |\sigma |$.
	Since $|\alpha | \leq |\sigma '|$ and $\alpha$, $\sigma '$ are prefixes of $\chi_A$, we have
	$\alpha \sqsubseteq \sigma '$.
	Define $\beta$ to be $\sigma '$.

	For the other direction, let $X$ be $\cla$  loc-meager as witnessed by $h$, i.e.
	for every $A \in X$ there exists $i \in \N$ such that $h_i$ avoids $A$. Let $N$ be a Turing machine
	computing $h$. Let $f: \N \rightarrow \N$ be as in Lemma \ref{l:lt2}, and let $Q$ be a deterministic Turing machine
	computing $f$.
	We define a $\Delta$-loc-computable  constructor $g$ 
	inducing a winning strategy for player II in the game
	$G[X,\N^{\N} , \Delta\text{-}\loc]$. 
	We show that for any strategy $s$, $R(s,g)$ meets $h_i$ for every
	$i\in \N$, which implies  $R(s,g) \not \in X$.
	Here is a description of a Turing machine $M$ computing $g$.
	For a string $\sigma$ with $n=\log |\sigma |$ , $M^{\sigma}(s_{|\sigma |},k)$  does the following.
	\begin{enumerate}
		\item 	Compute $$B= \max_{m\geq 1} [f(m) \leq n]$$ in the following way.
			For $t=1,2,\ldots$ compute $f(t)$ by simulating
			$Q$ for  $n^2$ steps, and denote the result by $b_t$, until either $Q$ doesn't
			halt, or $b_t >n$. Let $t_0$ denote the first $t$ for which this happens,
			then define $B=b_{t_0 -1}$.

	 	\item 	\label{i:unb}Compute $n_0 = \min \{t: t\leq B, \text{ and }(\forall \tau \sqsubseteq \sigma \mbox{ such that }
			|\tau |\leq B) \ \ h_t (\tau) \not \sqsubseteq \sigma \}$.

		\item	If no such $n_0$ exists output $0$ if $k=1$, and output $\perp$ if $k>1$.

		\item	\label{i:troisb}If $n_0$ exists, then if $k=1$ output 0, otherwise simulate
			$N^{\sigma 0}(s_{|\sigma| +1},n_0,k-1)$ answering $N$'s
			queries in $\dom(\sigma  0) \backslash \dom(\sigma )$ with $0$, and output the result of the simulation.
	\end{enumerate}

	Let us check that $g$ is  $\Delta$-loc-computable. For the first step, we have that
	whenever $f(m)\leq n$, $Q$ halts within $O(n)$ steps. Since  $Q$ is simulated
	$n^2$ steps, $B$ is  computed correctly, in polynomial time.
	For the second step, the $B^3 $ simulations of $N$  can be done in time  $u\in\Delta$.
	Moreover every $h_{t}(\tau)$ computed during the second step has size at most $B$, thus
	only the first $n$ bits of the input $\sigma$ need to be read. This together with the fourth step
	guarantees that the query set for $M$ is given by
	$$G_M(|s_{|\sigma| }|,k)= \{ 1,2,\ldots  n \} \cup G_N(|s_{|\sigma | +1}|,n, k-1)$$
	which has
	size $u'(\log |\sigma| )$, for some $u'\in\Delta$.

	We show that $R(s,g)$ meets every $h_i$ for any strategy $s$. Indeed suppose this is not the case, i.e.
	there is a strategy  $s$ such that $R(s,g)$ does not meet $h$.
	Let $n_0$ be the smallest index such that $R(s,g)$ does not meet $h_{n_0}$.
	Since $R(s,g)$ meets $h_{n_0 -1}$ there is a string $\tau$ such that
	$$h_{n_0 -1}(\tau) \sqsubset R(s,g)\ .$$
	Since $g$ strictly extends strings at every round, after a certain number of  rounds,
	$s$  outputs a string $\sigma$ long enough to enable step two (of $M$'s description) to
	find out that
	$$h_{n_0 -1}(\tau) \sqsubseteq \sigma$$
	thus incrementing $n_0 -1$ to $n_0$.
	At this round we have
	$$g(\sigma)= \sigma 0 \, \ext (h_{n_0}(\sigma  0))$$
	i.e.
	$$h_{n_0} \sqsubset R(s,g)$$ which is a contradiction.
	\qed
	\end{proof}
	
	\subsection{Local Categories on BPP}

        In this section we introduce local categories on the probabilistic class
        \bpp .
        To this end we need the following probabilistic strategies.

	\begin{definition}  \label{d:bpplog}
		An  indexed   strategy $h: \N \times \bool{*} \rightarrow \bool{*}$
		is called $\bpp$-loc-computable
		if
		there is a probabilistic random access Turing machine (as defined in Definition \ref{d:pcompindexedstratgy})
		$M$ such that for every $\tau \in \bool{*}$
		and every $i,k,n \in \N$,
		$$\Pr [M^{\tau}(s_{|\tau|}, i,k,n)=\ext (h_i (\tau),k)] \geq 1-2^{-n}$$
		where the probability is taken over the internal coin tosses of $M$,
		$M$ runs in time polynomial in $\log |\tau| + |i| +|k|+n$,
		and there is a poly printable query set $G$ such that for every
		$m,i,k \in \N$ and for every $i',k' \in \N$ (such that $i'\leq i$ and
		$k'\leq k$), and for every input $\sigma \in \bool{*}$  (such that  $\log |\sigma| \leq m$),
		$M^{\sigma}(s_{|\sigma|},i,k,n)$ queries $\sigma$ only on bits that
		are in $G(m,i,k)$; where $G(m,i,k)$ is printable in time polynomial
		in $m+|i|+|k|$.
	\end{definition}

	\begin{remark}
	By using standard Chernoff bound arguments it is easy to show that Definition
	\ref{d:bpplog} is robust, i.e. the error probability can range from
	$\frac{1}{2} +\frac{1}{p(n)}$ to $1-2^{-q(n)}$ for any polynomials $p,q$, without enlarging or reducing
	the class of strategies defined this way.
	\end{remark}

        Similarly to the deterministic case, a class $X$ is called meager if there is a single probabilistic strategy that avoids
        $X$.

	\begin{definition}
 		A class of languages $X$ is $\bpp$-loc-meager if there exists a $\bpp$-loc-computable indexed
		strategy $h$, such that for every $L \in X$ there exists $i\in \N$, such that
		$h_i$ avoids $L$.
	\end{definition}

        The definition of enumerable infinite unions is similar to Definition \ref{d:finiteunions}.

        Let us prove that all three basic properties hold for locally-computable Baire categories  on \bpp .
	\begin{theorem}\label{t:ax1bpp}
		For any language L  in $\bpp$,  $\{ L\} $ is $\bpp$-loc-meager.
	\end{theorem}
	\begin{proof}
	The proof is similar to Theorem  \ref{t:ax1ploc} except that the constructor $h$ is computed with
	error probability smaller than $2^{-n}$.
       	\qed
	\end{proof}

        The second basic property is easy to show.
	\begin{theorem}\label{t:ax2bpp}
			A $\bpp$-loc-union of $\bpp$-loc-meager sets is $\bpp$-loc-meager.
	\end{theorem}
	\begin{proof}
		Similar to Theorem \ref{t:ax2p}.
	\qed
	\end{proof}

        Let us prove the third basic property.
	\begin{theorem}

		$\bpp$ is not $\bpp$-loc-meager.

	\end{theorem}
	\begin{proof}

	The proof is similar to Theorem \ref{t:ax3ploc} except for the second step of $N$'s computation,
	where every simulation of $M$ is performed with error probability smaller than $2^{-n}$.
	Since  there are $n$ distinct simulations of $M$, the total error probability
	is  smaller than $n2^{-n}$, which ensures that $L$ is in \bpp .
	\qed
	\end{proof}


\subsection{SPARSE is Meager in P}

        It was shown in Section \ref{c.category}  that the class of languages with subexponential density is not meager for the  first category notion
        in this paper (the language in the proof of Theorem \ref{t:ax3p} has subexponential density).
        Here we prove that local computable strategies are stronger than the strategies of Section~\ref{c.category},
        by showing that the class \sparse\ (and also the class of languages of subexponential density) is $\p$-loc-meager. 
	The idea of the proof is to extend any prefix of a language with 
	enough ones to make sure it is not sparse.

	\begin{theorem}
		\sparse\ is  $\p$-loc-meager.
	\end{theorem}
	\begin{proof}
	Let $L$ be any sparse language. Then there exists a polynomial $p$ such that
	$$|L\cap \bool{n}| \leq p(n) \text{ for every } n \geq 1. $$
	Consider the following strategy $h$, which on input
	$\sigma\in\bool{*}$ pads $\sigma$ with $|\sigma |$ $1$'s.
	Since $L$ is sparse,  $h$ avoids $L$.
	We construct a random access Turing machine $M$ for $h$;
	on input  $\sigma \in \bool{*}$ and $j\in\N$,
	$M^{\sigma}(s_{|\sigma| },j)$ outputs 1 if $1\leq j\leq |\sigma |$
	and $\perp$ otherwise. Since $M$ doesn't query its oracle, $h$ is
	$\p$-loc-computable which ends the proof.
	\qed
	\end{proof}


\subsection{Meager-Comeager Laws} \label{s.all-meager-pspace}

	The following meager-comeager laws in $\pspace$ and $\bpp$ contrast with the resource-bounded measure case, where many of 
	those all or nothing laws are not known to hold.
	\begin{theorem}
		Let $X\in \{ \zpp ,\rp , \bpp , \np \} $. Then either $X$ is $\cla$-loc-meager or $X=\cla$.
	\end{theorem}
	\begin{proof}
		We need the following lemma, whose proof is an extension of a similar result in \cite{Fenner:rb_local_category}.
	
	\begin{lemma} \label{t:weak-tech}
         Let $X$ be a $\Sigma^0_2$ class, such that there exists a language $A$ in $\cla$,
         such that for every finite variant $A'$ of $A$, $A' \not \in X$.
         Then $X$ is $\cla$  loc-meager.
      \end{lemma}
	Let us show the lemma.
	The idea of the proof is to extend any prefix of a language according to language $A$ until
	until we know we have avoided $X$.

      By hypothesis there exists a polynomial oracle  Turing machine $M$, such that
      $$X=\{ L| \exists x \forall y: M^L(x,y)=0\} \ .$$
      Consider the following $\cla$  loc-computable strategy $g$ where $\ext(g(\sigma) , k))$,
      with $n=\log |\sigma|$, is computed  as follows.

      \begin{enumerate}
         \item   Simulate $M^L(x,y)$ for every $x<\log n$ and $y <\log k$, where
                 $$\chi_L = \sigma A(s_{|\sigma| +1}) A(s_{|\sigma |+2})\cdots$$
         \item   If for every $x<\log n$ there exists  $y <\log k$ such that
                 $M^L(x,y) \neq 0$ output $\perp$, else output $A(s_{|\sigma| +k})$.
      \end{enumerate}

      Let us show that $g$ is a strategy.
      Suppose for a contradiction that $g$ extends $\sigma$ infinitely.
      Then the result is a finite variant of $A$, hence not in $C$.
      Therefore there exists $k\in \N$, such that
      $$(\forall x < |s_{|\sigma|}|)  (\exists y < \log k)M^L(x,y)\neq 0$$
      where
      $L=g(\sigma)$.
      Hence $g$ should not have extended $\sigma$ more than $k$ bits, which is a contradiction.

      $g$ is $\Delta$-loc-computable    since there are  $\log n \cdot \log k$
      simulations to perform and since the queries  to $A$ can be computed in $t$  steps ($t\in\Delta$).
      The query set $G_g(n,k)$ has size $t(n+|k|)$ for some $t\in \Delta$, therefore $g$
      is $\Delta$-loc-computable.

      Let us show that $g$ avoids $X$.
      Denote by $L$ the result of the game between $g$ and some strategy $f$, where player II
      plays according to $g$.
      Let $z$ be any string.
      On the first turn for player II where the state of the game is of length
      at least $2^{2^{z+1}}$, player II  extends ensuring that
      $$(\forall x < z+1)( \exists y < \log k) M^L (x,y)\neq 0\ .$$
      Thus
      $$M^L(z,y) \neq 0$$
      which implies $L\not\in X$.

      This ends the proof of the lemma.
      We have the following consequences.

      \begin{corollary}  \label{c:sig}
         Let $X$ be a $\Sigma^0_2$ class closed under finite variants. Then
         $X$ is $\cla$-loc-meager  iff $\cla \not \subseteq  X$.
      \end{corollary}
      Which ends the proof of the theorem.
 	\end{proof}

	Similarly meager-comeager laws in \bpp\ can be proved.

	\begin{theorem}
		Let $X\in \{\p , \zpp ,\rp  \} $. Then either $X$ is $\bpp$-loc-meager or $X=\bpp$.
	\end{theorem}
	\begin{proof}
		It is easy to check that Lemma  \ref{t:weak-tech} also holds in  \bpp .
	\qed
	\end{proof}

\subsection{Weak Completeness}\label{s:weak-completeness}

        The concept of weak completeness was introduced in \cite{Lutz:weaklyhard-sets}.
        A set $A$ is called $\cla$-weakly-complete
        if its lower span (the class of sets reducible to $A$) does not have  $\cla$-measure zero.
        Lutz showed in \cite{Lutz:weaklyhard-sets} the existence of $\expc$-weakly-complete sets that are not $\expc$-complete.
        Similarly we can define a categorical weak completeness notion, by calling a set $A$
        $\cla$-loc-weakly-complete if its lower span is not $\cla$-loc-meager.
        We show that there is no $\p$-loc-weakly-complete  incomplete language,   i.e.
        $\p$-loc-weakly-completeness is equivalent to $\p$-completeness.

      \begin{theorem}
         $\p$-loc-weakly-completeness is equivalent to $\p$-completeness, under Turing logspace reductions.
      \end{theorem}
	\begin{proof}
      Let $A\in\p$ be any language.
      It is easy to check that the lower span $A^{\geq^{\log}_T}$ is a  $\Sigma^0_2$ class and is closed under finite variants.
      Thus by Corollary \ref{c:sig} we have  $A^{\geq^{\log}_T}$ is not $\p$-loc-meager iff $A$ is
      $\leq^{\log}_{T}$-hard for \p .
      \qed
	\end{proof}

      Another consequence of Corollary   \ref{c:sig}
      is the meagerness of the class of complete sets for \p , under the assumption
      \p\ is not equal to $\dspace (\log n)$.
      \begin{theorem}  \label{t:weakc}
          If \p\ is not equal to $\dspace (\log n)$, then
	   the class of $\leq^{\log}_{T}$-$\p$-complete sets   is $\p$-loc-meager.
      \end{theorem}
	\begin{proof}
      Let $A$ be a $\leq^{\log}_{T}$-$\p$-complete set.
      Consider $A^{\leq^{\log}_T}$ the upper span of $A$.
      It is easy to check that
      $A^{\leq^{\log}_T}$ verifies the hypothesis of Corollary \ref{c:sig}.
      By our assumption, $A$ cannot reduce to a set in $\dspace (\log n)$, so $\p \not \subseteq A^{\leq^{\log}_T}$,
		hence
      $A^{\leq^{\log}_T}$ is $\p$-loc-meager. Since every $\leq_{T}^{\log}$-$\p$-complete language 
      is in $A^{\leq^{\log}_T}$, this ends the proof.
      \qed
      \end{proof}

	Note that the same result holds unconditionally for locally-computable categories  on $\qplin$.
      	\begin{theorem}
	   	The class of $\leq_{T}^{\log}$-$\qplin$-complete sets  is $\qplin$-loc-meager.
	\end{theorem}
	\begin{proof}
	The proof is similar to Theorem \ref{t:weakc}.
	\end{proof}



\subsection{Measure vs Baire Categories}

	Although Lemma \ref{t:weak-tech} shows that locally computable strategies are extremely strong,
	the following easy observation shows that the size notion yielded from resource-bounded measure
	is incomparable  with  the one derived from Baire category. This might explain why many
	locally computable Baire category results  on small complexity classes 
	(meager-comeager laws in \pspace\ and \bpp , equivalence between \p -loc-weak-completeness and \p -completeness, 
	conditional smallness of the class of \p -complete languages) 
	are not known to hold in the resource-bounded
	measure setting on small complexity classes.

	We use the measure notion on \p\ of \cite{Moser:dim-in-P}. We shall give a brief description of it.
	A martingale is a function $d:\{ 0,1 \}^{*} \rightarrow \Rplus$ such that,
	for every $w \in \{ 0,1 \}^{*}$,
	$$2d(w) = d(w0) + d(w1).$$ 
	A martingale $d$ succeeds on language $L$ if $$\limsup_{n\rightarrow\infty} (L[1\ldots n])=\infty . $$
	Informally speaking, a set has $\p$-measure zero if there exists a martingale $d$ 
	computable in polynomial time that succeeds on every language in the class. 
	For full details we refer the reader to \cite{Moser:dim-in-P}.
	A language $R$ is $\p$-random if every polynomial time martingale does not succeed on $R$.  
	A language $G$ is $\p$-loc-generic if every $\p$-loc-computable strategy does not avoid $G$.

	The following result shows that $\p$-measure and $\p$-loc-categories are incomparable.
	The idea of the proof is that for any $\p$-random language it is impossible that 
	the language contains no strings of size $n$ for infinitely many lengths $n$; such a language can then 
	easily be avoided by a locally-computable strategy. The other direction is due to the existence of 
	$\p$-loc-generic languages containing significantly more zeroes that ones in the limit, which makes it possible
	for a martingale to succeed on them.

	\begin{theorem}
	\text{ }
		\begin{enumerate}
			\item	Every \p -random set is $\p$-loc-meager.
			\item	There exist $\p$-loc-generic sets which have \p- measure zero.
		\end{enumerate}
	\end{theorem}
	\begin{proof}
		Let $R$ be \p -random, then
		$$\forallio n: L^{=n}_{n}\neq \emptyset \label{e:property}$$
		(where $L^{=n}_{n}$ denotes the $n$ first strings of size $n$)
		otherwise consider the following
		\p -computable martingale $d$ which divides its initial capital into shares $c_n=1/n^2$,
		and uses  capital $c_n$ to bet on strings of size $n$, by betting all the  capital that
		the first $n$ strings of size $n$ have membership bit 0.
		Whenever this bet is correct for strings of size $n$,
		$d$ wins $2^n/n^2$. Since
		$$\existsio n: L_{n}^{=n}\neq \emptyset$$
		$d$'s capital grows unbounded
		on $R$, which contradicts $R$'s \p -randomness.
		Consider the following $\p$-loc-computable strategy $h$.
		By hypothesis there exists a constant $k$ such that
		$$\forall n > k: L^{=n}_{n}\neq \emptyset \ .$$
		Therefore strategy $h$ defined by
		$$\ext (h(\sigma ))= 0^{2|\sigma |+2^{2k}}$$
		avoids $R$, i.e. $\{R\}$ is $\p$-loc-meager.

		For the second part consider the following language $G$.
		Let $M_i$ be an (non-effective) enumeration of all Turing machines computing $\p$-loc-computable
		strategies. Consider the following characteristic sequence of $G$ .
		$$\chi_G = \und{1}_{B_0} \ \und{ \ext (h_1(B_0))}_{B_1} \ \und{0 \cdots 0}_{B_2} \
		\und{ \ext (h_2(B_0  B_1  B_2))}_{B_3} \ \und{0 \cdots 0}_{B_4} \cdots
		$$
		where block $B_{2i}$ contains $5\cdot |B_0 B_1 \cdots B_{2i-1}|$ $0$'s.
		By construction $G$ is $\p$-loc-generic because it meets every $\p$-loc-computable strategy.
		Consider the same \p -computable martingale $d$ as above.
		By construction of $G$, the zones padded with $0$'s are large enough to guarantee that
		$$\existsio n: G^{=n}_{n}= \emptyset \ .$$
		Therefore $d$'s capital grows unbounded on $G$
		which ends the proof.
		\qed
	\end{proof}
	
	\section{Conclusion}
	We have introduced two Baire category notions on small deterministic and probabilistic classes,
	and given applications of both notions in derandomization, circuit complexity, meager-comeager laws
	and weak-completeness, some of which are not known to hold with respect to measure in small complexity 
	classes. We then observed that categories and measure on small classes are incomparable, which might 
	explain these differences between the two settings.


\textbf{Acknowledgments}\\
We thank the anonymous referees for their  useful comments.


\end{document}